\documentclass{book}    
\usepackage{piers}  
\begin{document}

\title{Adjoint charge in electromagnetic field}
\maketitle

\author      {Zi-Hua Weng}
\affiliation {Xiamen University}
\address     {}
\city        {Xiamen}
\postalcode  {}
\country     {China}
\phone       {345566}    
\fax         {233445}    
\email       {xmuwzh@xmu.edu.cn}  
\misc        { }  
\nomakeauthor

\begin{authors}

{\bf Zi-Hua Weng}\\
\medskip
School of Physics and Mechanical \& Electrical Engineering,
\\Xiamen University, Xiamen 361005, China\\

\end{authors}

\begin{paper}

\begin{piersabstract}
Making use of the octonion operator, the electromagnetic field
generates an adjoint field theoretically. The source of adjoint
field includes the adjoint charge and the adjoint current. The
adjoint charge has an impact on the gravitational mass and the mass
distribution in the electromagnetic field with its adjoint field,
and causes further the predictions to departure slightly from the
conservation of mass. The inferences can explain why the adjoint
charge will influence the mass distribution in the gravitational
field and electromagnetic field of celestial bodies. And then the
adjoint charge can be considered as one kind of candidate for the
dark matter.
\end{piersabstract}


\psection{Introduction}

The mass and the 'missing mass' both are crucial physical
conceptions for various field theories. There is only the conception
of mass in the existing electromagnetic field and gravitational
field, which are described with the vectorial quantity. But this
perspective can not explain why there is the 'missing mass'
\cite{zwicky, bournaud} in the universe up to now.

The algebra of quaternions \cite{adler} was first used by J. C.
Maxwell to describe the electromagnetic field. While the octonions
\cite{cayley} can be used to demonstrate the characteristics of
gravitational field and electromagnetic field, including the
conservation of mass etc. The related theoretical inferences are
only dealt with the quaternion operator but the octonion operator
\cite{weng}. In the octonion space, the operator should be extended
from the quaternion operator to the octonion operator.

Making use of the octonion operator, the electromagnetic field
demonstrated by the octonions will generate an adjoint field. The
source of adjoint field includes the adjoint charge and adjoint
current. The adjoint charge and its movement can not be observed by
usual experiments. However, when the adjoint charge is combined with
the ordinary charge to become the charged particles, their movements
will be accompanied by some mechanical or electric effects.

The electromagnetic field and its adjoint field both can be
demonstrated by the quaternions, although they are quite different
from each other indeed. With the property of octonions, we find that
the adjoint charge has an influence on the conservation of mass in
the electromagnetic field. The adjoint charge takes part in the
gravitational interaction and impacts the mass distribution, and
then can be considered as one kind of candidate for dark matter
\cite{krasnov, diemand}.

\psection{Octonion transformation}

The quaternion spaces can be used to describe the electromagnetic
field or the gravitational field. Meanwhile, two quaternion spaces
can combine together to become an octonion space. The latter  can be
used to demonstrate the electromagnetic field and gravitational
field simultaneously.

The quaternion space can be considered as the two-dimensional
complex space, and the octonion space as the two-dimensional
quaternion space. In the quaternion space for the gravitational
field, the basis vector is $\mathbb{E}_g$ = ($1$,
$\emph{\textbf{i}}_1$, $\emph{\textbf{i}}_2$,
$\emph{\textbf{i}}_3$), and the radius vector is $\mathbb{R}_g$ =
($r_0$, $r_1$, $r_2$, $r_3$), with the velocity $\mathbb{V}_g$ =
($v_0$, $v_1$, $v_2$, $v_3$). For the electromagnetic field, the
basis vector is $\mathbb{E}_e$ = ($\emph{\textbf{I}}_0$,
$\emph{\textbf{I}}_1$, $\emph{\textbf{I}}_2$,
$\emph{\textbf{I}}_3$), the radius vector is $\mathbb{R}_e$ =
($R_0$, $R_1$, $R_2$, $R_3$), with the velocity $\mathbb{V}_e$ =
($V_0$, $V_1$, $V_2$, $V_3$).

The $\mathbb{E}_e$ is independent of the $\mathbb{E}_g$, with
$\mathbb{E}_e$ = $\mathbb{E}_g$ $\circ$ $\emph{\textbf{I}}_0$ . The
$\circ$ denotes the octonion multiplication. The $\mathbb{E}_g$ and
$\mathbb{E}_e$ can be combined together to become the basis vector
$\mathbb{E}$ of the octonion space.
\begin{eqnarray}
\mathbb{E} = \mathbb{E}_g + \mathbb{E}_e = (1, \emph{\textbf{i}}_1,
\emph{\textbf{i}}_2, \emph{\textbf{i}}_3, \emph{\textbf{I}}_0,
\emph{\textbf{I}}_1, \emph{\textbf{I}}_2, \emph{\textbf{I}}_3)
\end{eqnarray}

In the octonion space, the radius vector $\mathbb{R}$ is
\begin{eqnarray}
\mathbb{R} = \Sigma ( r_i \emph{\textbf{i}}_i ) + \Sigma ( R_i
\emph{\textbf{I}}_i)~,
\end{eqnarray}
and the velocity $\mathbb{V}$  is
\begin{eqnarray}
\mathbb{V} = \Sigma ( v_i \emph{\textbf{i}}_i ) + \Sigma ( V_i
\emph{\textbf{I}}_i)~.
\end{eqnarray}
where, $r_0 = v_0 t$; $v_0$ is the speed of light, $t$ is the time.
$i = 0, 1, 2, 3$, $j = 1, 2, 3$, $\emph{\textbf{i}}_0 = 1$.

In some special cases, one ordinary mass $m$ can be combined with
one ordinary charge $q$ to become an ordinary particle, such as the
proton and electron etc. And then, we can measure their various
characteristics, and have following relation.
\begin{eqnarray}
R_i \emph{\textbf{I}}_i = r_i \emph{\textbf{i}}_i \circ
\emph{\textbf{I}}_0~;~ V_i \emph{\textbf{I}}_i = v_i
\emph{\textbf{i}}_i \circ \emph{\textbf{I}}_0~.
\end{eqnarray}

The octonion quantity $\mathbb{D} (d_0, d_1, d_2, d_3, D_0, D_1,
D_2, D_3 )$ is defined as follows.
\begin{eqnarray}
\mathbb{D} = d_0 + \Sigma (d_j \emph{\textbf{i}}_j) + \Sigma (D_i
\emph{\textbf{I}}_i)
\end{eqnarray}
where, $d_i$ and $D_i$ are all real.

When the coordinate system is transformed into the other, the
physical quantity $\mathbb{D}$ will be transformed into one new
octonion $\mathbb{D}' (d'_0 , d'_1 , d'_2 , d'_3 , D'_0 , D'_1 ,
D'_2 , D'_3 )$ .
\begin{eqnarray}
\mathbb{D}' = \mathbb{K}^* \circ \mathbb{D} \circ \mathbb{K}
\end{eqnarray}
where, $\mathbb{K}$ is one octonion, and $\mathbb{K}^* \circ
\mathbb{K} = 1$; $*$ denotes the conjugate of octonion.

If the $d_0$ does not take part in the above transformation, it
satisfies the following relation.
\begin{eqnarray}
d_0 = d'_0
\end{eqnarray}

In the above equation, the scalar part $d_0$ is preserved during the
octonion coordinates are transforming. Some scalar invariants of
electromagnetic field will be obtained from this characteristics of
the octonion.

\begin{table}[h]
\caption{The octonion multiplication table.}
\label{tab:table1}
\centering
\begin{tabular}{ccccccccc}
\hline \hline
$ $ & $1$ & $\emph{\textbf{i}}_1$  &
$\emph{\textbf{i}}_2$ & $\emph{\textbf{i}}_3$  &
$\emph{\textbf{I}}_0$  & $\emph{\textbf{I}}_1$
& $\emph{\textbf{I}}_2$  & $\emph{\textbf{I}}_3$  \\
\hline $1$ & $1$ & $\emph{\textbf{i}}_1$  & $\emph{\textbf{i}}_2$ &
$\emph{\textbf{i}}_3$  & $\emph{\textbf{I}}_0$  &
$\emph{\textbf{I}}_1$
& $\emph{\textbf{I}}_2$  & $\emph{\textbf{I}}_3$  \\
$\emph{\textbf{i}}_1$ & $\emph{\textbf{i}}_1$ & $-1$ &
$\emph{\textbf{i}}_3$  & $-\emph{\textbf{i}}_2$ &
$\emph{\textbf{I}}_1$
& $-\emph{\textbf{I}}_0$ & $-\emph{\textbf{I}}_3$ & $\emph{\textbf{I}}_2$  \\
$\emph{\textbf{i}}_2$ & $\emph{\textbf{i}}_2$ &
$-\emph{\textbf{i}}_3$ & $-1$ & $\emph{\textbf{i}}_1$  &
$\emph{\textbf{I}}_2$  & $\emph{\textbf{I}}_3$
& $-\emph{\textbf{I}}_0$ & $-\emph{\textbf{I}}_1$ \\
$\emph{\textbf{i}}_3$ & $\emph{\textbf{i}}_3$ &
$\emph{\textbf{i}}_2$ & $-\emph{\textbf{i}}_1$ & $-1$ &
$\emph{\textbf{I}}_3$  & $-\emph{\textbf{I}}_2$
& $\emph{\textbf{I}}_1$  & $-\emph{\textbf{I}}_0$ \\
\hline $\emph{\textbf{I}}_0$ & $\emph{\textbf{I}}_0$ &
$-\emph{\textbf{I}}_1$ & $-\emph{\textbf{I}}_2$ &
$-\emph{\textbf{I}}_3$ & $-1$ & $\emph{\textbf{i}}_1$
& $\emph{\textbf{i}}_2$  & $\emph{\textbf{i}}_3$  \\
$\emph{\textbf{I}}_1$ & $\emph{\textbf{I}}_1$ &
$\emph{\textbf{I}}_0$ & $-\emph{\textbf{I}}_3$ &
$\emph{\textbf{I}}_2$  & $-\emph{\textbf{i}}_1$
& $-1$ & $-\emph{\textbf{i}}_3$ & $\emph{\textbf{i}}_2$  \\
$\emph{\textbf{I}}_2$ & $\emph{\textbf{I}}_2$ &
$\emph{\textbf{I}}_3$ & $\emph{\textbf{I}}_0$  &
$-\emph{\textbf{I}}_1$ & $-\emph{\textbf{i}}_2$
& $\emph{\textbf{i}}_3$  & $-1$ & $-\emph{\textbf{i}}_1$ \\
$\emph{\textbf{I}}_3$ & $\emph{\textbf{I}}_3$ &
$-\emph{\textbf{I}}_2$ & $\emph{\textbf{I}}_1$  &
$\emph{\textbf{I}}_0$  & $-\emph{\textbf{i}}_3$
& $-\emph{\textbf{i}}_2$ & $\emph{\textbf{i}}_1$  & $-1$ \\
\hline
\end{tabular}
\end{table}

\psection{Electromagnetic field}

By means of the octonion operator, the electromagnetic field will
generate an adjoint field in the octonion space. The adjoint field
is derived from the electromagnetic field potential. The source of
adjoint field includes the adjoint charge and adjoint electric
current. In case of the electromagnetic field is accompanied by the
adjoint field, the adjoint charge has an influence on the
gravitational mass, and then causes some mechanical or electric
effects. As a result, the adjoint charge may be considered as one
kind of candidate for the dark matter.

The electromagnetic field potential is
\begin{eqnarray}
\mathbb{A}_e = \Sigma (A_i \emph{\textbf{I}}_i)~.
\end{eqnarray}

In the electromagnetic field, the field strength $\mathbb{B}_e =
\Sigma (b_{ei} \emph{\textbf{i}}_i) + \Sigma (B_{ei}
\emph{\textbf{I}}_i)$ consists of the electromagnetic strength
$\mathbb{B}_{eg}$ and adjoint strength $\mathbb{B}_{ee}$ .
\begin{eqnarray}
\mathbb{B}_e = \lozenge \circ \mathbb{A}_e = \mathbb{B}_{eg} +
\mathbb{B}_{ee}
\end{eqnarray}
where, $\mathbb{B}_{ee} = \Sigma (b_{ei} \emph{\textbf{i}}_i)$,
$\mathbb{B}_{eg} = \Sigma (B_{ei} \emph{\textbf{I}}_i)$; $ \lozenge
= \Sigma \emph{\textbf{i}}_i ( \partial/\partial r_i) + \Sigma
\emph{\textbf{I}}_i ( \partial/\partial R_i)$.

In the above, we choose the following gauge conditions to simplify
succeeding calculation.
\begin{eqnarray}
\partial A_0 / \partial r_0 - \Sigma (\partial A_j /
\partial r_j) = 0 ~,~
\partial A_0 / \partial R_0 + \Sigma (\partial A_j /
\partial R_j) = 0 ~.
\end{eqnarray}

The adjoint field strength $\mathbb{B}_{ee}$ in Eq.(9) includes two
components, $\textbf{g}_e = ( g_{e01} , g_{e02} , g_{e03} ) $ and
$\textbf{b}_e = ( g_{e23} , g_{e31} , g_{e12} )$ .
\begin{eqnarray}
\textbf{g}_e/v_0 = \emph{\textbf{i}}_1 ( \partial_0 A_1 -
\partial_1 A_0 ) + \emph{\textbf{i}}_2 ( \partial_0 A_2 - \partial_2
A_0 ) + \emph{\textbf{i}}_3 ( \partial_0 A_3 - \partial_3 A_0 )
\\
\textbf{b}_e = \emph{\textbf{i}}_1 ( \partial_3 A_2 -
\partial_2 A_3 ) + \emph{\textbf{i}}_2 ( \partial_1 A_3 - \partial_3
A_1 ) + \emph{\textbf{i}}_3 ( \partial_2 A_1 - \partial_1 A_2 )
\end{eqnarray}

Meanwhile, the electromagnetic field strength $\mathbb{B}_{eg}$
involves two parts, $\textbf{E}_e = ( B_{e01} , B_{e02} , B_{e03} )
$ and $\textbf{B}_e = ( B_{e23} , B_{e31} , B_{e12} )$ .
\begin{eqnarray}
\textbf{E}_e/v_0 = \emph{\textbf{I}}_1 ( \partial_0 A_1 +
\partial_1 A_0 ) + \emph{\textbf{I}}_2 ( \partial_0 A_2 + \partial_2
A_0 ) + \emph{\textbf{I}}_3 ( \partial_0 A_3 + \partial_3 A_0 )
\\
\textbf{B}_e = \emph{\textbf{I}}_1 ( \partial_3 A_2 -
\partial_2 A_3 ) + \emph{\textbf{I}}_2 ( \partial_1 A_3 - \partial_3
A_1 ) + \emph{\textbf{I}}_3 ( \partial_2 A_1 - \partial_1 A_2 )
\end{eqnarray}

The electric current density $\mathbb{S}_{eg} = q \mathbb{V}_g \circ
\emph{\textbf{I}}_0$ is the source of electromagnetic field, and its
adjoint current density $\mathbb{S}_{ee} = \bar{q} \mathbb{V}_g $ is
that of adjoint field. And they can be combined together to become
the field source $\mathbb{S}_e$ . In the octonion space, the
electromagnetic source $\mathbb{S}_e$ can be defined from the
electromagnetic field strength  $\mathbb{B} = k_b \mathbb{B}_e$ .
\begin{eqnarray}
\mu \mathbb{S} = - ( \mathbb{B}/v_0 + \lozenge)^* \circ \mathbb{B} =
k_b (\mu_{ee} \mathbb{S}_{ee} + \mu_{eg} \mathbb{S}_{eg}) -
\mathbb{B}^* \circ \mathbb{B}/v_0
\end{eqnarray}
where, $k_b^2 = \mu_{gg} /\mu_{eg}$; $\mu_{gg}$, $\mu_{ee}$, and
$\mu_{eg}$ are the coefficients. $\bar{q}$ is the adjoint charge.

The $\mathbb{B}^* \circ \mathbb{B}/(2\mu_{gg})$ is the field energy
density.
\begin{eqnarray}
\mathbb{B}^* \circ \mathbb{B}/ \mu_{gg} = \mathbb{B}_e^* \circ
\mathbb{B}_e / \mu_{eg}
\nonumber
\end{eqnarray}

The above means that the electromagnetic field and its adjoint field
both make a contribution to the gravitational mass in the octonion
space.

\psection{Conservation of mass}

In the electromagnetic field and its adjoint field, for one charged
particle with inertial massless, the linear momentum density
$\mathbb{P} = \mu \mathbb{S} / \mu_{gg}$ is written as
\begin{eqnarray}
\mathbb{P} = \widehat{m} v_0 + \Sigma ( M_q V_i \emph{\textbf{i}}_i
\circ \emph{\textbf{I}}_0 ) + \Sigma ( M_e v_i \emph{\textbf{i}}_i
)~.
\end{eqnarray}
where, $\widehat{m} = - (\mathbb{B}^* \circ \mathbb{B} / \mu_{gg}
)/v_0^2 $ ; $M_q = q k_b \mu_{eg}/\mu_{gg}$ ; $M_e = \bar{q} k_b
\mu_{ee}/\mu_{gg}$ .

The above means that the gravitational mass density $(\widehat{m} +
M_e)$ is changed with all kinds of field strengthes in the
electromagnetic field and its adjoint field. From Eq.(6), we have
one linear momentum density, $\mathbb{P}' (p'_0, p'_1, p'_2, p'_3,
P'_0, P'_1, P'_2, P'_3)$, when the octonion coordinate system is
rotated. And we obtain the invariant equation from Eqs.(7) and (16).
\begin{eqnarray}
(\widehat{m} + M_e ) v_0 = (\widehat{m}' + M'_e) v'_0
\end{eqnarray}

Under Eqs.(3), (7), and (17), we find the gravitational mass density
$(\widehat{m} + M_e)$ remains unchanged.
\begin{eqnarray}
\widehat{m} + M_e = \widehat{m}' + M'_e
\end{eqnarray}

The above means that the gravitational mass density $(\widehat{m} +
M_e)$ will keep unchanged, under the octonion coordinate
transformation in Eq.(6) in the electromagnetic field and its
adjoint field.

\psection{Conclusion}

In the octonion space, the electromagnetic field described by the
octonion operator will generate an adjoint field. In some cases, the
electromagnetic field will be accompanied by its adjoint field. And
that the source of adjoint field will impact the mass distribution
and the conservation of mass in the electromagnetic field and
gravitational field, especially in the universe.

In the electromagnetic field with its adjoint field, the adjoint
field exerts an influence on the gravity in two aspects. On the one
hand, the adjoint charge presents to the quaternion space for the
gravitational field, and is one part of the gravitational mass. The
adjoint charge is similar to the mass, and possesses the
characteristics of gravity. On the other hand, the gravitational
mass density is changed with the electromagnetic strength and
adjoint field strength. The gravitational mass takes part in the
gravitational interaction, so that the adjoint field strength will
effect the gravity. It states that the conservation of mass will be
changed with the adjoint field strength and adjoint charge.
Therefore the adjoint charge can be considered as one kind of dark
matter in the astronomy related to the electromagnetic field.

It should be noted that the study of adjoint charge in the
electromagnetic field and its adjoint field examined only one simple
case with very weak field strength in the electromagnetic field with
its adjoint field. Despite its preliminary character, this study can
clearly indicate the field strength of electromagnetic field and its
adjoint field have an influence on the gravity and the conservation
of mass. For the future studies, the related investigation will
concentrate on only the predictions of the gravity fluctuation and
the mass distribution, in the strong adjoint field strength of
electromagnetic field with its adjoint field.

\ack
This project was supported partially by the National Natural
Science Foundation of China under grant number 60677039.

\end{paper}

\end{document}